\documentclass[submission, Phys]{SciPost}
\usepackage{amsfonts,amsmath,amssymb}
\usepackage{mathrsfs,frcursive}
\usepackage{color}
\usepackage{theorem,cite}
\usepackage{hyperref}
\newcommand{\bea}{\begin{eqnarray}}
\newcommand{\eea}{\end{eqnarray}}
\newcommand{\beali}{\begin{align}}
\newcommand{\eeali}{\end{align}}
\newcommand{\beano}{\begin{eqnarray*}}
\newcommand{\eeano}{\end{eqnarray*}}
\newcommand{\beq}{\begin{equation}}
\newcommand{\eeq}{\end{equation}}

\newcommand{\mb}[1]{\hspace{2.1ex}\mbox{#1}\hspace{2.1ex}}

\def\fo{{\mathfrak o}}
\def\fp{{\mathfrak p}}

\def\fr{{\mathfrak r}}
\def\fs{{\mathfrak s}}
\def\ft{{\mathfrak t}}

  \def\cH{{\cal H}}  
    
    \def\cO{{\cal O}}
    
\def\cS{{\cal S}}

\newcommand{\CC}{{\mathbb C}}

\newcommand{\II}{{\mathbb I}}

\newcommand{\RR}{{\mathbb R}}

\newcommand{\ZZ}{{\mathbb Z}}

\newcommand{\mR}{{\mathscr R}}

\newcommand{\mL}{{\mathscr L}}

\def\sr{\mathsf{r}}

\newcommand{\br}{\boldsymbol{\sr}}

\newcommand{\wh}[1]{\widehat{#1}}

\newcommand{\half}{{\textstyle{\frac{1}{2}}}}

\def\sgn{\mathop{\rm sgn}\nolimits}

\newcommand{\so}{\scriptscriptstyle \rm I}
\newcommand{\st}{\scriptscriptstyle \rm I\hspace{-1pt}I}
\newcommand{\sth}{\scriptscriptstyle \rm I\hspace{-1pt}I\hspace{-1pt}I}

\numberwithin{equation}{section}

\begin{document}
%%%%%%%%%%%%%%%%%%%%%%%%%%%%%%%%%%%%%%%%%%%%%%%%%%%%%%%%%%%%%%%
\pagestyle{empty}

\null
\vspace{20pt}

\parskip=6pt

\begin{center}
\begin{LARGE}
\textbf{A new  integrable structure associated to
\\[1ex] 
the Camassa--Holm peakons}
\end{LARGE}
\end{center}

\begin{center}
J. Avan\textsuperscript{(a)}, 
L. Frappat\textsuperscript{(b)}, 
E. Ragoucy\textsuperscript{(b)*} 
\footnote{avan@cyu.fr, luc.frappat@lapth.cnrs.fr, eric.ragoucy@lapth.cnrs.fr}
\end{center}

\begin{center}
{\bf (a)} Laboratoire de Physique Th\'eorique et Mod\'elisation, \\ 
CY Cergy Paris Universit\'e, CNRS, F-95302 Cergy-Pontoise, France
\\
{\bf (b)} Laboratoire d'Annecy-le-Vieux de Physique Th{\'e}orique LAPTh, \\ 
CNRS, Université Savoie Mont Blanc, F-74940 Annecy
\\
\end{center}

\begin{center}
\today
\end{center}

\vspace{4mm}

\section*{Abstract}
{\bf
We provide a closed Poisson algebra involving the Ragnisco--Bruschi generalization of peakon dynamics in the Camassa--Holm shallow-water equation. This algebra is generated by three independent matrices. From this presentation, we propose a one-parameter integrable extension of their  structure. It leads to  a new $N$-body peakon solution to the Camassa--Holm shallow-water equation depending on two parameters.
\\
We  present two explicit constructions of a (non-dynamical) $r$-matrix formulation for this new Poisson algebra. The first one relies on a tensorization of the $N$-dimensional auxiliary space by a 4-dimensional space. We identify a family of Poisson commuting quantities in this framework, including the original ones. This leads us to constructing a second formulation identified as a spectral parameter representation.
}

\vspace{10pt}
\noindent\rule{\textwidth}{1pt}
\tableofcontents
\thispagestyle{fancy}
\noindent\rule{\textwidth}{1pt}
\vspace{10pt}

%%%%%%%%%%%%%%%%%%%%%%%%%%%%%%%%%%%%%%%%%%%%%%%%%%%%%%%%%%%%%%%
%%%%%%%%%%%%%%%%%%%%%%%%%%%%%%%%%%%%%%%%%%%%%%%%%%%%%%%%%%%%%%%

\section{Introduction\label{sec:intro}}
Many non-linear two-dimensional $(x,t)$ integrable fluid equations exhibit so-called peakon
 solutions which take the generic form
\begin{equation}
\label{eq:peakonphi}
u(x,t) = \sum_{i=1}^N p_i(t)\, e^{-|x-q_i(t)|}\,.
\end{equation}
Their dynamics for $(p_i,q_i)$ is deduced from a reduction of the 1+1 fluid equations for $u(x,t)$ and may have integrability properties.
The best known example of such integrability feature for peakon equations is given by the Camassa--Holm shallow-water equation \cite{CH,CHH}
\begin{equation}\label{eq:CH0}
u_t - u_{xxt} + 3 uu_x = 2u_x u_{xx} + uu_{xxx}\,.
\end{equation}
Integrability of the peakons themselves was studied in particular in \cite{dGHH}. A one-parameter extension was proposed by Ragnisco and Bruschi \cite{RB}, who proved integrability from an implicit construction of a dynamical  $r$-matrix.
An explicit construction of this dynamical $r$-matrix is still lacking. In this paper, we propose an alternative construction of the Poisson structure relevant for integrability. It relies on a non-dynamical $r$-matrix formulation and uses three dynamical Lax-type matrices, instead of the two matrices introduced by Ragnisco and Bruschi. A direct consequence is that the new integrable peakon model depends on two coupling constants. It provides a more general peakon dynamics which in turns yields a new peakon-type solution of the
Camassa--Holm  equation \eqref{eq:CH0}.

We present two representations of the non-dynamical $r$-matrix formulation. The first one requires the use of a larger auxiliary space obtained by tensoring the initial $N$-dimensional space by a 4-dimensional space. One advantage of this representation is that the Yang--Baxter equation for the $r$-matrix structure takes a remarkably simple and compact form. The second representation involves the introduction of a spectral parameter in the $N$-dimensional auxiliary space. Both formulations provide an algebraic framework to construct the same hierarchy of conserved quantities.

The plan of the paper runs as follows. 

Section \ref{sec:PB} describes the complete Poisson algebra of the three relevant matrix generators
involved in our formulation of the integrability properties. It allows the construction of a family of Poisson commuting quantities, including a two-parameter generalization of the peakon Hamiltonian. From the $N$-body dynamics triggered by this new Hamiltonian we deduce a new peakon solution of the Camassa--Holm equation. 

Section \ref{sec:rmat} is devoted to the construction of the $r$-matrix in the extended auxiliary space picture. The explicit computation of the classical Yang--Baxter equation is performed. It is in fact a modified classical Yang--Baxter equation, where the r.h.s.  is built from the three-fold tensored Casimir operator of the algebra.

Finally, Section \ref{sec:cons} displays the construction of Poisson commuting quantities. It is first done in the framework of the extended auxiliary space. The existence of families of Poisson commuting quantities naturally induces the existence of an alternative, spectral parameter presentation.

\section{Poisson structure\label{sec:PB}}
\subsection{Description of the original model and its first generalisation \label{sec:Lax}}
The $n$-peakon solutions \eqref{eq:peakonphi} of the Camassa--Holm equation
yield a dynamical system for $p_i,q_i$:
\begin{equation}\label{eq:pqdot}
\dot{q}_i = \sum_{j=1}^N p_j \, e^{-|q_i-q_j|} \,, \qquad \dot{p}_i = \sum_{j=1}^N p_ip_j \,\fs_{ij}\, e^{-|q_i-q_j|} \,,
\end{equation}
where $\fs_{ij}=\sgn({q}_i-{q}_j)$.
This discrete dynamical system is described by a Hamiltonian
\begin{equation}\label{eq:Hpeak}
H_{CH} = \frac12\,\sum_{i,j=1}^N {p_ip_j} \,e^{ -| q_i-q_j| }
\end{equation}
 such that
\begin{equation}
\dot{f}=\{ f\,,\,H_{CH} \} \,,
\end{equation}
with the canonical Poisson structure:
\begin{equation}\label{eq:PB-can}
\{ p_i,p_j \} = \{ q_i,q_j \} = 0 \,, \qquad \{ q_i,p_j \} = \delta_{ij} \,.
\end{equation}
The dynamics is encoded in the Lax formulation \cite{CF,CH}
\begin{equation}
\frac{dL}{dt} = [L,M]
\end{equation}
with
\begin{equation}
\label{eq:LCH}
L = \sum_{i,j=1}^N L_{ij} E_{ij} \,, \qquad L_{ij} = \sqrt{p_ip_j} \,e^{-\half | q_i-q_j | } \,,
\end{equation}
where $E_{ij}$ is the $N\times N$ elementary matrix with 1 at position $(i,j)$ and 0 elsewhere.
The Hamiltonian \eqref{eq:Hpeak} is recast as $H_{CH}=\frac12\text{Tr}L^2$.

A generalisation of this original integrable model was proposed by Ragnisco--Bruschi \cite{RB}, with a Lax matrix\footnote{One may suppose $\rho\in\CC$, but the corresponding Hamiltonians are not real anymore.}
 $L(\rho) = T + \rho S$, $\rho\in\RR$, where 
\begin{eqnarray}
T &=& \sum_{i,j} \sqrt{p_ip_j} \cosh \frac{\nu}{2}(q_i-q_j) \, E_{ij}\,,
\\
S &=& \sum_{i,j} \sqrt{p_ip_j} \sinh \frac{\nu}{2}|q_i-q_j| \, E_{ij}\,.
\end{eqnarray}
The proof of integrability relies on the construction of a dynamical $r$-matrix \cite{RB}.
The Hamiltonian takes the form
\begin{equation}\label{eq:HpeakRB}
\begin{split}
&H_{RB}(\rho) =\frac12\text{Tr}L(\rho)^2\\
&\qquad=\frac12 \,\sum_{i,j=1}^N {p_ip_j}\Big(\frac{\rho^2+1}{2} \cosh\big( {\nu}|q_i-q_j|\big)
+\rho\,\sinh\big( {\nu}|q_i-q_j|\big)-\frac{\rho^2-1}{2}\Big).
\end{split}
\end{equation}
One recovers the original Hamiltonian $H_{CH}$ for the values $\rho=-1$, $\nu=1$ or $\rho=1$, $\nu=-1$.

\subsection{General non dynamical Poisson Structure}
The description of the full algebraic structure associated to the Poisson brackets \eqref{eq:PB-can} and the Lax matrix $L(\rho)$ requires the introduction of a third matrix 
\begin{eqnarray}
A &= &\sum_{i,j} \sqrt{p_ip_j} \sinh \frac{\nu}{2}(q_i-q_j) \, E_{ij}
\end{eqnarray}
which allows to close the Poisson structure of $(T, S)$. It reads
\begin{align}
\big\{ T_1 , T_2 \big\} &= \frac{\nu}{8}\;\big[ \Pi+\Pi^{t},A_1-A_2 \big] 
\label{eq:PB1}\\
\big\{ A_1 , A_2 \big\} &= \frac{\nu}{8}\;\big[ \Pi-\Pi^{t},A_1-A_2 \big] \\
\big\{ S_1 , S_2 \big\} &= -\frac{\nu}{4}\;\big[ {\Gamma}_{12},S_1 \big] + \frac{\nu}{4}\;\big[ {\Gamma}_{21},S_2 \big] - \frac{\nu}{8}\;\big[ \Pi+\Pi^{t},A_1-A_2 \big] \\
\big\{ T_1 , A_2 \big\} &= \frac{\nu}{8}\;\left( \big[ \Pi-\Pi^{t},T_1 \big] - \big[ \Pi+\Pi^{t},T_2 \big] \right) \\
\big\{ A_1 , T_2 \big\} &= \frac{\nu}{8}\;\left( \big[ \Pi+\Pi^{t},T_1 \big] - \big[ \Pi-\Pi^{t},T_2 \big] \right) \\
\big\{ S_1 , T_2 \big\} &= \frac{\nu}{4}\;\big[ {\Gamma}_{21},T_2 \big] \qquad\qquad\quad\ 
\big\{ T_1 , S_2 \big\} = -\frac{\nu}{4}\;\big[ {\Gamma}_{12},T_1 \big] \\
\big\{ S_1 , A_2 \big\} &= \frac{\nu}{4}\;\big[ {\Gamma}_{21},A_2 \big] \qquad\qquad\qquad
\big\{ A_1 , S_2 \big\} = -\frac{\nu}{4}\;\big[ {\Gamma}_{12},A_1 \big] 
\label{eq:PB7}
\end{align}
with 
\begin{equation}\label{eq:defT}
\begin{split}
&\Pi=\sum_{i,j} E_{ij}\otimes E_{ji} \mb{;} \Pi^t=\sum_{i,j} E_{ij}\otimes E_{ij}\,,
\\
& {\Gamma}_{12}=\sum_{i>j}\Big( E_{ij}\otimes E_{ji} - E_{ji}\otimes E_{ij} +E_{ij}\otimes E_{ij} - E_{ji}\otimes E_{ji}\Big)\,.
\end{split}
\end{equation}
In \eqref{eq:PB1}--\eqref{eq:PB7}, we have used the auxiliary space description: for any $N\times N$ matrix $M$, we define 
$M_1=M\otimes \II_N$ and $M_2=\II_N\otimes M= \Pi\, M_1\,\Pi$. Similarly, for any matrix $M_{12}\in \text{End}(\CC^N)\otimes \text{End}(\CC^N)$, we define 
$M_{21}=\Pi\,M_{12}\,\Pi$.

Remark that, following \cite{BBT}, ${\Gamma}_{12}$ can be identified with a classical $r$-matrix for the classical open Toda chain.

\subsection{Generalized peakons}
From the above Poisson structure, it is natural to introduce the most general Lax matrix 
\begin{equation}\label{eq:defLbar}
\bar L(\rho,\lambda) = T + \rho S + \lambda A\,, 
\end{equation}
hereafter denoted $ \bar L$.
The Poisson structure of this Lax matrix reads
\begin{equation}\label{eq:PBLbar}
\begin{aligned}
\big\{ \bar L_1 , \bar L_2 \big\} =& -\frac{\nu}{4} \Big( \rho\big[ {\Gamma}_{12},\bar L_1 \big] - \rho\big[ {\Gamma}_{21},\bar L_2 \big] - \lambda \big[ \Pi-\Pi^{t},\bar L_1-\bar L_2 \big] + \lambda\rho \big[ \Pi,S_1-S_2 \big]  \\
& \qquad - \lambda^2 \big[ \Pi-\Pi^{t},A_2 \big] - (1-\rho^2)\big[ \Pi+\Pi^{t},A_1 \big] \Big)\,.
\end{aligned}
\end{equation}
We  ask the traces $\tau_n=\text{Tr}(\bar L^n)$ to be Poisson commuting for all values of $n\in\ZZ_+$. A direct calculation leads to 
\begin{equation}
\begin{aligned}
\big\{ \tau_n , \tau_m \big\} =&\sum_{i=0}^{n-1}\sum_{j=0}^{m-1}\text{Tr}_{12}(\bar L_1)^{i} (\bar L_2)^{j}\big\{ \bar L_1 , \bar L_2 \big\} (\bar L_1)^{n-1-i} (\bar L_2)^{m-1-j}
\\
=&nm\,\text{Tr}_{12}(\bar L_1)^{n-1} (\bar L_2)^{m-1}\big\{ \bar L_1 , \bar L_2 \big\} \,.
\end{aligned}
\end{equation}
We first notice that for any matrix $M$ we have 
\begin{equation}\label{eq:trPi}
\begin{aligned}
&\text{Tr}_{12}(\bar L_1)^{n-1} (\bar L_2)^{m-1}\big[ \Pi , M_1 \big] = \text{Tr}_{12}(\bar L_1)^{n-1} (\bar L_2)^{m-1}\big( \Pi  M_1 -M_1\Pi\big)
\\
&\qquad\qquad
=\text{Tr}_{12}\Big((\bar L_1)^{n-1} \Pi(\bar L_1)^{m-1}   M_1 -(\bar L_1)^{n-1}M_1 \Pi(\bar L_1)^{m-1}\Big)
\\
&\qquad\qquad=\text{Tr}_{1}\Big((\bar L_1)^{n-1} (\bar L_1)^{m-1}   M_1 -(\bar L_1)^{n-1}M_1 (\bar L_1)^{m-1}\Big)
\\
&\qquad\qquad=0\,,
\end{aligned}
\end{equation}
where we have used that for any matrices $U$, $V$, $U_2\,V_1=V_1\,U_2$  and $U_2\,\Pi=\Pi\,U_1$ (to get the second line),
 the property $\text{Tr}_{2}\Pi=\II_{N}$ (third line) and the cyclicity of the trace (fourth line).
Similarly, we have $\text{Tr}_{12}(\bar L_1)^{n-1} (\bar L_2)^{m-1}\big[ \Pi , M_2 \big]=0$, so that 
when computing $\{ \tau_n , \tau_m \}$,
the terms corresponding to $\Pi$ in \eqref{eq:PBLbar} can be dropped.
Thanks to this property, we get
\begin{equation}
\begin{aligned}
\big\{ \tau_n , \tau_m \big\} =& -\frac{nm\nu}{4} \text{Tr}_{12}(\bar L_1)^{n-1} (\bar L_2)^{m-1}\Big(
\big[r_{12}\,,\,\bar L_1\big] -\big[r_{21}\,,\, \bar L_2\big] \Big)\\
&+\frac{nm\nu}{4} \text{Tr}_{12}(\bar L_1)^{n-1} (\bar L_2)^{m-1}\,\left((1- \rho^2) \big[ \Pi^t\,,\,  A_1 \big] - \lambda^2 \big[ \Pi^t\,,\,  A_2 \big] \right)
 \,,
\end{aligned}
\end{equation}
where $r_{12}=\rho\, \Gamma_{12}+\lambda\, \Pi^t$. 

Now,
for any matrix $M$ we have 
 $\big[ \Pi^t,M_2 \big]=\big[ \Pi^t,M_1^t\big]$ and since $A$ is an  antisymmetric matrix, we get
\begin{equation}\label{eq:PBtau}
\begin{aligned}
\big\{ \tau_n , \tau_m \big\} =& -\frac{nm\nu}{4} \text{Tr}_{12}(\bar L_1)^{n-1} (\bar L_2)^{m-1}\Big(
\big[r_{12}\,,\,\bar L_1\big] -\big[r_{21}\,,\, \bar L_2\big] -(1- \rho^2+\lambda^2) \big[ \Pi^t\,,\,  A_1 \big]\Big) \,.
\end{aligned}
\end{equation}
Furthermore, starting from the relations
\begin{equation}\label{eq:trR=0}
\begin{split}
&\text{Tr}_{12}(\bar L_1)^{n-1} (\bar L_2)^{m-1}
\big[R_{12}\,,\,\bar L_1\big] =\frac1n\,\text{Tr}_{2}\Big((\bar L_2)^{m-1}
\text{Tr}_{1}\big[R_{12}\,,\,\bar L_1^n\big]\Big)=0\,,\\
&\text{Tr}_{12}(\bar L_1)^{n-1} (\bar L_2)^{m-1}\big[R_{21}\,,\, \bar L_2\big]=0\,,
\end{split}
\end{equation}
valid for any matrix $R_{12}$, we get when $R_{12}=\Pi^t=R_{21}$
\begin{equation}\label{eq:ruseJ}
\text{Tr}_{12}(\bar L_1)^{n-1} (\bar L_2)^{m-1}
\big[\Pi^t\,,\,\bar L_1-\bar L_2\big]=0\,.
\end{equation}
Finally, using that 
\begin{equation}
\begin{split}
&\big[\Pi^t\,,\, M_1- M_2\big]=0\qquad\qquad\text{for any symmetric matrix $M$},\\ 
&\big[\Pi^t\,,\, U_1- U_2\big]=2\big[\Pi^t\,,\, U_1\big]\qquad\text{for any anti-symmetric matrix $U$}, 
\end{split}
\end{equation}
we deduce that $\big[\Pi^t\,,\,\bar L_1-\bar L_2\big]=2\lambda\,\big[\Pi^t\,,\, A_1\big]$, so that from \eqref{eq:ruseJ}  
we get 
\begin{equation}
\text{Tr}_{12}(\bar L_1)^{n-1} (\bar L_2)^{m-1}\big[\Pi^t\,,\, A_1\big]=0\,.
\end{equation}
Then, \eqref{eq:PBtau} rewrites as
\begin{equation}
\begin{aligned}
\big\{ \tau_n , \tau_m \big\} =& -\frac{nm\nu}{4} \text{Tr}_{12}(\bar L_1)^{n-1} (\bar L_2)^{m-1}\Big(
\big[r_{12}\,,\,\bar L_1\big] -\big[r_{21}\,,\, \bar L_2\big]\Big)=0 \,.
\end{aligned}
\end{equation}
Thus, the model associated to $\bar L$ defines an integrable double deformation of the original peakon model.

The Hamiltonian corresponding to these new peakons reads
\begin{equation}\label{eq:Hpeak_new}
\begin{split}
&H_{new}(\rho,\lambda) =\frac12\text{Tr}\bar L^2= 
H_{RB}(\rho)+\frac{1}{2}\lambda^2\,\sum_{i,j=1}^N {p_ip_j} \sinh^2 \big(\frac{\nu}{2}|q_i-q_j|\big)\\
&\qquad=\frac12\,\sum_{i,j=1}^N {p_ip_j}\Big(\frac{\rho^2-\lambda^2+1}{2} \cosh\big( {\nu}|q_i-q_j|\big)
+\rho\,\sinh\big( {\nu}|q_i-q_j|\big)-\frac{\rho^2-\lambda^2-1}{2}\Big)\\
&\qquad=\sum_{i,j=1}^N {p_ip_j}\Big(\frac{(\rho+1)^2-\lambda^2}{4}\,e^{ \nu\,| q_i-q_j| } +\frac{(\rho-1)^2-\lambda^2}{4}\,e^{ -\nu\,| q_i-q_j| } \Big)
-\frac{\rho^2-\lambda^2-1}{4}\,\fp^2\,,
\end{split}
\end{equation}
where we introduced 
\begin{equation}\label{eq:defP}
\fp=\sum_{i=1}^Np_i\,.
\end{equation}
We obtain $H_{CH}$ for $\lambda=0$, $\rho=-1$ and $\nu=1$, while
 $H_{RB}(\rho)$ is recovered when $\lambda=0$. 
 
 Note also that for $(\rho+1)^2=\lambda^2$ and $\nu=1$, we get a shifted version of the original peakon model:
 \begin{equation}
H_{new}\big(\rho,\pm(\rho+1)\big) \Big|_{\nu=1}= -\rho\,H_{CH} -\frac{\rho+1}2\,\fp^2\,.
\end{equation}
A similar result holds for  $(\rho-1)^2=\lambda^2$ and $\nu=-1$.

Remark that a generic value for $\nu$ can be obtained from the cases $\nu=\pm1$, see section \ref{sec:fluid} below.
Hence the above conditions correspond to a condition on $\rho$ and $\lambda$, rather than two conditions on 
$\rho$, $\lambda$ and $\nu$.

\subsection{$N$-body solutions of the fluid equation\label{sec:fluid}}
  We establish here a general result on the consistency conditions for peakon-type $N$-body solutions to (a deformation of) the  Camassa--Holm equation, including the Hamiltonian evolution of the $N$-body variables. 
We first consider the case $\nu^2=1$, and then show how a generic value for $\nu$ can be obtained from the cases $\nu=\pm1$. 
\subsubsection*{A deformed version of the Camassa--Holm equation}
We first restrict ourself to the case
  \begin{equation}
  \nu^2=1\,.
  \end{equation}
The Hamiltonian $H_{new}(\rho,\lambda)$ given in \eqref{eq:Hpeak_new}, can be rewritten as
\begin{equation}\label{eq:Hnew}
H = \frac12 \sum_{i,j=1}^N p_ip_j F(|q_i-q_j|) \,,
\end{equation}
with 
\begin{equation}\label{eq:soluF}
 F(q)= \frac{\rho^2-\lambda^2+1}{2} \cosh\big( {\nu}q\big)
+\rho\,\sinh\big( {\nu}q\big)-\frac{\rho^2-\lambda^2-1}{2}\,.
\end{equation}
Note that the function $F$ in \eqref{eq:soluF} obey the following differential equation with the initial value conditions
\begin{equation}\label{eq:eqdiff-F}
 F''(x)-F(x)=\frac{\rho^2-\lambda^2-1}{2}\,,\qquad F(0)=1\,,\qquad F'(0)=\nu\,\rho\,,
\end{equation}
where we have used the property $\nu^2=1$.

 The Hamiltonian $H_{new}(\rho,\lambda)$ describes a  time evolution for  $p_i,q_i$, given  by:
\begin{equation}\label{eq:mvt-new}
\begin{split}
\dot{q}_i &= \big\{q_i\,,\,H_{new}(\rho,\lambda)\big\}=\sum_{j=1}^N p_j \, F(|q_i-q_j|)\,,\\
\dot{p}_i &= \big\{p_i\,,\,H_{new}(\rho,\lambda)\big\}=\, \sum_{j=1}^N p_ip_j \,\fs_{ij}\, F'(|q_i-q_j|)\,,
\end{split}
\end{equation}
where $\fs_{ij}=\sgn({q}_i-{q}_j)$ as above, and 
\begin{equation}
F'(q)=\frac{d}{dq}F(q)={\nu}\Big(\frac{\rho^2-\lambda^2+1}{2} \sinh\big( {\nu}q\big)
+\rho\,\cosh\big( {\nu}q\big) \Big)\,.
\end{equation}

To recover the time evolutions \eqref{eq:mvt-new} from a fluid equation, we define $u(x,t)$ as  
\begin{equation}\label{eq:peakonphi-new}
u(x,t) = \sum_{i=1}^N p_i(t)\, F({-|x-q_i(t)|})\,
\end{equation}
which is a direct generalisation of  \eqref{eq:peakonphi}.
Plugging the form \eqref{eq:peakonphi-new} into the l.h.s. of the differential relation \eqref{eq:eqdiff-F}
and using \eqref{eq:defP},
one finds 
\begin{equation}
\begin{split}
&u_t - u_{xxt} + 3 uu_x - 2u_x u_{xx} - uu_{xxx} = \\
&
=\sum_{i=1}^N\Big\{ \mu\,\dot{p_i} 
+\sgn(x-q_i)p_i\Big(\dot{q_i}\,(F'_i-\,F'''_i)+2\mu\,\fp\, F_i'\Big)
\\
&\quad +2\,F'(0)\,\delta(x-q_i)\,\big(\dot{p_i}-\sum_{j=1}^N\fs_{ij}p_ip_jF'_{ij}  \big)
+2\,F'(0)\,\delta'(x-q_i)\,\big(-\dot{q_i}+\sum_{j=1}^Np_jF_{ij}  \big)p_i\Big\}\,,
\end{split}
\end{equation}
with
 \begin{equation}
\mu=\,\frac{\rho^2-\lambda^2-1}{2}\,.
\end{equation}

We have introduced the notation 
\begin{equation}
F_i=F(-|x-q_i|)\ \text{ and }\ F_{ij}=F(-|q_i-q_j|).
\end{equation}

{From} the differential equations \eqref{eq:eqdiff-F} and the equations of motion \eqref{eq:mvt-new}, we get 
\begin{equation}
\begin{split}
u_t - u_{xxt} + 3 uu_x - 2u_x u_{xx} - uu_{xxx} &= 2\mu\,\fp\,\sum_{i=1}^N\,\sgn(x-q_i)p_i\, F_i'\,.
\end{split}
\end{equation}
To obtain this relation, we have used the property (deduced from the equations of motion \eqref{eq:mvt-new}) that $\fp$ defined in \eqref{eq:defP} is a free constant parameter of the model, i.e.
$\dot{\fp} =0.$

Now remarking that $\sum_{i=1}^N\,\sgn(x-q_i)p_iF_i'=u_x$, we find a modification of the  Camassa--Holm shallow-water equation
\begin{equation}
\label{eq:CH-defo}
 u_t - u_{xxt} + 3\, uu_x - 2\,u_x u_{xx} - uu_{xxx} =2\mu\,\fp\,u_x\,.
\end{equation}
For $\mu=0$ we recover the undeformed Camassa--Holm shallow-water equation. For $\mu\neq 0$, we find the deformed fluid equation \eqref{eq:CH-defo}. It exhibits peakon-type solutions \eqref{eq:peakonphi-new} with the integrable dynamics \eqref{eq:Hpeak_new}.

\subsubsection*{Deformed versus undeformed Camassa--Holm equation}
We now show that this deformed equation is in fact equivalent to the original one.
We perform the following change of variables and function:
\begin{equation}\label{eq:change}
y=x-\mu\,\fp\,t\,,\quad t'=t,\quad v=u-\mu\,\fp\,.
\end{equation}
Starting from the equation \eqref{eq:CH-defo}, we get back to the undeformed Camassa--Holm  equation: 
\begin{equation}
\label{eq:CH-new}
 v_t - v_{yyt} + 3\, vv_y - 2\,v_yv_{yy} - vv_{yyy} =0\,.
\end{equation}

As a consequence, the expression \eqref{eq:soluF} for $F$ provides a new $N$-body solution of the Camassa--Holm equation:
\begin{equation}
u(x,t) =-\mu\,\fp+\sum_{i=1}^N p_i(t)\, F\Big({-|x+\mu\,\fp\, t-q_i(t)|}\Big)\,.
\end{equation}
In fact, since $\fp$ is a free constant of motion, we get a one-parameter family of solutions.

\subsubsection*{Rescaling by $\nu$}
We wish to show that the case $\nu>0$ can be obtained from the case $\nu=1$. 
We start with the model defined with the value $\nu=1$, with
\begin{equation}
\begin{split}
 u_{\nu=1}( x, t) &= -\mu\,\fp+\sum_{i=1}^N  p_i( t)\, F_{\nu=1}({-\,| x+\mu\,\fp\, t- q_i( t)|})\,,\\
 F_{\nu=1}(q)&= \frac{\rho^2-\lambda^2+1}{2} \cosh\big( q\big)+\rho\,\sinh\big( q\big)-\frac{\rho^2-\lambda^2-1}{2}\,,\\
H_{\nu=1} &= \frac12 \sum_{i,j=1}^N p_ip_j F_{\nu=1}(|q_i-q_j|) \,.
\end{split}
\end{equation}
We perform the symplectic transformation
$\bar q_i=\frac1{\nu}\,q_i$,  $\bar p_i={\nu}\,p_i$, and at the same time a dilation $\bar x=\frac1{\nu}\,x$.
It is easy to see that the Hamiltonian $H_{\nu=1}$ get a $\frac1{\nu^2}$ factor, indicating that we need to define 
$\bar t=\frac1{\nu^2}\,t$. Then 
\begin{equation}\label{eq:ubar}
\bar u(\bar x,\bar t) = -\mu\,\bar \fp+\sum_{i=1}^N  \bar p_i( t)\, F_{\nu=1}({-\nu\,|\bar x+\mu\,\bar\fp\, \bar t-\bar q_i(\bar t)|})\,
\end{equation} 
obeys 
\begin{equation}
\nu^2\,(\bar u_{\bar t} + 3\, \bar u\bar u_{\bar x}) = \bar u_{\bar x\bar x\bar t} + 2\bar u_{\bar x} \bar u_{\bar x\bar x} + \bar u\bar u_{\bar x\bar x\bar x}\,.
\end{equation}
In \eqref{eq:ubar}, $\bar q_i$ and  $\bar p_i$ are canonical variables whose time evolution is triggered by the Hamiltonian 
$\bar H=\frac1{\nu^2}\,H_{\nu=1}$.

Note that to obtain \eqref{eq:ubar}, we have used the property $|\nu|=\nu$, hence the condition $\nu>0$.
To get the other values of $\nu$, one needs to start from the model with $\nu=-1$ and perform the same transformations.

\section{$\mathscr{R}$-matrix representation\label{sec:rmat}}
\subsection{Extended auxiliary space}
Up to now, we have used $N$-dimensional auxiliary spaces denoted by 1 and 2. These spaces will be now labelled 0 and $0'$  respectively. We introduce two additional 4-dimensional auxiliary spaces labelled 1 and 2, and define the resulting tensored auxiliary spaces {\scriptsize{$\rm I$}}=(0,1) and {\scriptsize{$\rm I\!\rm I$}}=(0',2).
We introduce the following $16N^2\times16N^2$ diagonal  $r$-matrix and $4N\times4N$ diagonal Lax matrix
\begin{eqnarray}
 \mR_{\so,\st}&=& \frac12(\Pi_{00'}-\Pi_{00'}^{t_0})\,U_{12} + \frac12(\Pi_{00'}+\Pi_{00'}^{t_0})\,V_{12} -{\Gamma}_{00'}\,W_{12}\,,
 \\
 \mL_{\so} &=& A_0\,\II_1 +T_0\,X_1 +S_0\,Y_1,
\end{eqnarray}
where
\begin{equation}
\begin{split}
&X= \text{diag}(1,-1,0,0) \ ,\qquad 
Y= \text{diag}(0,0,1,-1)\ , \qquad\
\\
&U_{12}= \frac{\nu}4\Big(\II\otimes\II +\frac12\big(X_1^2-Y_1^2 -X_2^2 +Y_2^2\big)\Big)\ ,\\
& V_{12}=\frac{\nu}4\big(X_1X_2-Y_1Y_2\big)\ ,\quad W_{12}=\frac{\nu}4Y_2.
\end{split}
\end{equation}
In the following we will also write
\begin{equation}\label{eq:defR-L}
 \mR= \sum_{i,j=1}^4 \fr^{ij}_{00'}\, e_{ii}\otimes e_{jj} \quad
 \text{ and }\quad
 \mL = \sum_{i=1}^4 \ell^{i}_0\,e_{ii}\,,
\end{equation}
where $e_{ij}$ are the $4\times4$  elementary matrices. 
In \eqref{eq:defR-L}, the superscripts indicate the matrix entries in the $4\times4$ auxiliary spaces, 
while the subscripts show in which $N\times N$ auxiliary space they act on.
From the above expressions, we get
\begin{equation}\label{eq:rij-li}
\begin{split}
& \ell^{1}_{0}=A_0+T_0\ ,\qquad \ell^{2}_{0}=A_0-T_0\ ,\qquad \ell^{3}_{0}=A_0+S_0\ ,\qquad\ell^{4}_{0}=A_0-S_0\ ,\\
&\fr^{11}_{00'}=\fr^{22}_{00'}=\frac{\nu}4\Pi_{00'}\ ,\qquad\,\fr^{12}_{00'}=\fr^{21}_{00'}=-\frac{\nu}4\Pi^{t_0}_{00'}\ ,\\ 
&\fr^{13}_{00'}=\fr^{23}_{00'}=\frac{\nu}4(\Pi_{00'}-\Pi^{t_0}_{00'}-{\Gamma}_{00'})\ ,\qquad\fr^{14}_{00'}=\fr^{24}_{00'}=\frac{\nu}4(\Pi_{00'}-\Pi^{t_0}_{00'}+{\Gamma}_{00'})\ ,\\
&\fr^{33}_{00'}=-\frac{\nu}4(\Pi^{t_0}_{00'}+{\Gamma}_{00'})\ ,\qquad\fr^{34}_{00'}=\frac{\nu}4(\Pi_{00'}+{\Gamma}_{00'})\ ,\\
&\fr^{43}_{00'}=\frac{\nu}4(\Pi_{00'}-{\Gamma}_{00'})\ ,\qquad\fr^{44}_{00'}=\frac{\nu}4(-\Pi^{t_0}_{00'}+{\Gamma}_{00'})\ ,\\
&\fr^{31}_{00'}=\fr^{32}_{00'}=\fr^{41}_{00'}=\fr^{42}_{00'}=0\ .
\end{split}
\end{equation}

\subsection{Poisson brackets and classical Yang--Baxter relation}
The relation
\begin{equation}
\big\{ \mL_{\so}\, ,\, \mL_{\st} \big\}= \big[ \mR_{\so,\st}\, , \,\mL_{\so} \big] - \big[ \mR_{\st,\so}\, ,\, \mL_{\st} \big]
\end{equation}
is equivalent to
\begin{equation}
\big\{\ell^{i}_{0}\, ,\, \ell^{j}_{0'}\big\}= \big[ \fr^{ij}_{00'}\,,\, \ell^{i}_{0} \big] - \big[ \fr^{ji}_{0'0} \,,\, \ell^{j}_{0'}\big]\,,\qquad 1\leq i\leq j\leq 4\,.
\end{equation}
Using the expressions \eqref{eq:rij-li}, one can check by a direct calculation that these relations are equivalent to the Poisson brackets \eqref{eq:PB1}--\eqref{eq:PB7}. 

We have also established that $ \mR$ obeys a modified Yang--Baxter relation
\begin{equation}
\big[ \mR_{\so,\st} , \mR_{\so,\sth} + \mR_{\st,\sth} \big] + \big[ \mR_{\sth,\st} , \mR_{\so,\sth} \big] = \cO_{\so,\st,\sth}
\end{equation}
which reads in component
\begin{equation}\label{eq:ybe-compo}
\big[ \fr_{00'}^{ij} ,  \fr_{00''}^{ik}+ \fr_{0'0''}^{jk} \big] + \big[ \fr_{0''0'}^{kj} , \fr_{00''}^{ik} \big]=\fo_{00'0''}^{ijk}\,.
\end{equation}
Using again the expressions \eqref{eq:rij-li}, we compute the r.h.s. of \eqref{eq:ybe-compo}.
To simplify its presentation, we introduce the  operators
\begin{equation}
\begin{split}
\Omega_{00'0''} &= \omega_{00'0''} -(\omega_{00'0''})^{t_0t_{0'}t_{0''}} 
\quad\text{with}\quad \omega_{00'0''}=\sum_{a,b,c} E_{ab}\otimes E_{bc}\otimes E_{ca}\,.
\end{split}
\end{equation}
Then, the only non-vanishing $\fo_{00'0''}^{ijk}$ are given by
\begin{equation}\label{eq:oijk}
\begin{aligned}
&\fo^{111}=\fo^{222}=\fo^{333}=\fo^{444}=\Omega\,,\quad&\fo^{122}=\fo^{211}=\fo^{344}=\fo^{433}=-\Omega^{t_{0}}\,,\\
& \fo^{112}=\fo^{221}= \fo^{334}=\fo^{433}=-\Omega^{t_{0''}}\,,\quad 
& \fo^{121}=\fo^{212}=\fo^{343}=\fo^{434}=-\Omega^{t_{0'}} \,,
\end{aligned}
\end{equation}
where we omitted the subscript ${}_{00'0''}$ to lighten the writing.

To obtain \eqref{eq:oijk}, we have used the classical Yang--Baxter equation for $\Gamma$ \cite{AFR}:
\begin{equation}
[\Gamma_{00'},\Gamma_{00''}+\Gamma_{0'0''}]+[\Gamma_{0''0'}\,,\,\Gamma_{00''}]=\Omega-\Omega^{t_{0}}+\Omega^{t_{0'}}+\Omega^{t_{0''}}\,.
\end{equation}
Remark that the formulas \eqref{eq:oijk} imply the following formula for $\cO_{\so,\st,\sth}$:
\begin{equation}
\cO_{\so,\st,\sth}=\Omega_{0,0',0''}\, \cH_{123} -\sum_{s=1}^3 \Omega_{0,0',0''}^{\wh t_s}\,Ad(Z_s)(\cH_{123})
\,,\ \text{with}\ 
\begin{cases}\displaystyle \cH_{123}=\sum_{j=1}^4 e_{jj}\otimes e_{jj}\otimes e_{jj},\\
Z=e_{12}+ e_{21}+ e_{34}+ e_{43}.
 \end{cases}
\end{equation}
We have defined $\wh t_1=t_0$, $\wh t_2=t_{0'}$, $\wh t_3=t_{0''}$. The notation $Z_s$ stands for the matrix $Z$ acting in the auxiliary space $s$.

\section{Conserved quantities \label{sec:cons}}
\subsection{Conserved quantities from the $\mL$ operator}
Now that we have established a Lax presentation of the Poisson brackets, we can consider the traces 
\begin{equation}
\begin{split}
\ft^K_n&=\text{Tr}\,\text{tr}\big({ K}\,\mL^n\big)\\
&=k^{1}\,\text{Tr}[(A+T)^n] +k^{2}\,\text{Tr}[(A-T)^n] 
+k^{3}\,\text{Tr}[(A+S)^n] +k^{4}\,\text{Tr}[(A-S)^n]
\end{split}
\end{equation}
where ${K}=\sum_{i} k^i e_{ii} \otimes \II_N$ is a diagonal $4N\times4N$ matrix, acting as identity in the spaces $0$ and $0'$. 
We have denoted "Tr" the trace in the $N$-dimensional space and "tr" the trace in the $4$-dimensional space.
Since all matrices commute in the 4-dimensional space, it is easy to show that
\begin{equation}
\{\ft^K_n\,,\,\ft^{K'}_m\}=0\,,\quad \forall n,m\,.
\end{equation}
As a consequence we have a commuting family of operators generated by
\begin{equation}
\text{Tr}[(A+T)^n]\,,\quad\text{Tr}[(A-T)^n] 
\,,\quad\text{Tr}[(A+S)^n] \,,\quad\text{Tr}[(A-S)^n]\,.
\end{equation}
Since $A$ is antisymmetric while $T$ and $S$ are symmetric, we get using Tr$M=\text{Tr}(M^T)$:
\begin{equation}
\text{Tr}[(A+T)^n]=\text{Tr}[(T-A)^n] \quad\text{and}\quad 
\text{Tr}[(A+S)^n] =\text{Tr}[(S- A)^n]\,.
\end{equation}

To get more insight on the two remaining generators, we consider the first chamber
\begin{equation}
i<j\ \Leftrightarrow\ q_i<q_j\,.
\end{equation}
Then, looking at the form of $A$ and $S$, one sees that  
$A+S$ is a triangular matrix with zeros on the diagonal, so that
\begin{equation}\label{eq:traceSSbar}
\text{Tr}[(A+S)^n] =0\,.
\end{equation}
Similarly, a simple recursion shows that
\begin{equation}
(A+T)^n=\sum_{i_1,...,i_{n+1}} p_{i_2}\cdots p_{i_{n}}\,\sqrt{p_{i_1}p_{i_{n+1}}}\,e^{\frac{\nu}2(q_{i_1}-q_{i_{n+1}})}\, e_{i_1,i_{n+1}}\,.
\end{equation}
We then get
\begin{equation}\label{eq:traceST}
\text{Tr}[(A+T)^n] =\fp^n\,,
\end{equation}
where $\fp$ is given in \eqref{eq:defP}.

Since we obtain the other chambers through permutation of rows and columns, the properties \eqref{eq:traceSSbar}  and \eqref{eq:traceST} are valid everywhere. In conclusion, the quantities $\ft^K_n$ provide only one conserved quantity $\fp$ and its corresponding polynomial algebra.

\subsection{Adding more conserved quantities}
It is easy to show that 
 since $2T=\ell^1-\ell^2$, $2S=\ell^3-\ell^4$, and $2A=\ell^1+\ell^2=\ell^3+\ell^4$, we have
\begin{equation}
\begin{split}
\bar L
&=\frac12\Big( (1+\delta)\ell^1+ (-1+\delta)\ell^2 +(\rho+\lambda-\delta)\ell^3+(-\rho+\lambda-\delta)\ell^4\Big)\\
 &= \text{tr}({ D}\,\mL)\quad\text{with}\quad 
 D=\frac12\text{diag}\big(1+\delta\,,\,\delta-1\,,\,\lambda+\rho-\delta\,,\,\lambda-\rho-\delta\big)\,,
\end{split}
\end{equation}
which is valid for any $\delta\in\CC$. 
Then, one recovers the conserved quantities $\tau_n$ as 
\begin{equation}
\tau_n=\text{Tr}\Big(\big(\text{tr}\,D\,\mL\big)^n\Big)\,.
\end{equation}

Note that to prove this property, we need 
to consider $\big(\text{tr}D\mL\big)^n$ instead of $\mL^n$, which explains why the 
"natural" approach using the elements $\ft^K_n$ does not lead to the full set of conserved quantities.

Remark that 
 when $D$ is replaced by a general diagonal matrix $K$,  the set
\begin{equation}
\cS_K=\{\tau^K_n\,,\quad n=0,1,2...\}\quad \text{with}\quad
\tau^K_n=\text{Tr}\Big(\big(\text{tr}\,K\,\mL\big)^n\Big)\,
\end{equation}
is still commutative: $\{\tau_n^K\,,\,\tau_m^{K}\}=0$. 

We now consider two diagonal matrices $K$ and $K'$ and look for the vanishing of the Poisson brackets $\{\tau^{K'}_n\,,\,\tau_m^K\}$. 
Using a formal computation software, we are led to propose a  family of matrices 
\begin{equation}\label{eq:defK}
\begin{split}
&K(x)=diag\Big(-\frac{\rho-\lambda-1}{4\rho}\frac{\beta_1(x)}{x+1}\,,\,-\frac{\rho-\lambda+1}{4\rho}\frac{\beta_2(x)}{x+1}\,,\,
\frac{\beta_1(x)\beta_2(x)}{8\rho(x+1)}\,,\,0\Big)\\
&\beta_1(x)= (\rho+\lambda-1)x+3\rho-\lambda+1\quad\text{and}\quad \beta_2(x)= (\rho+\lambda+1)x+3\rho-\lambda-1
\end{split}
\end{equation}
depending on a parameter $x$, such that :
\begin{equation}\label{eq:PBtaux}
\{\tau_n(x)\,,\,\tau_m(y)\}=0\,,\quad \forall x,y\quad\text{with}\quad \tau_n(x)=\text{Tr}\Big(\big(\text{tr}K(x)\,\mL\big)^n\Big)\,.
\end{equation}
Note that there are other choices of $K$ matrices leading to the same hierarchy of conserved quantities $\tau_n(x)$.
The analytical proof of this property is given at the end of section \ref{sec:tau-a}. 
We first provide a spectral parameter formulation of the Poisson brackets.

\subsection{A spectral parameter representation of the Poisson structure\label{sec:spec}}
For fixed $\lambda$ and $\rho$, we introduce a family extending the $\bar L$ operator \eqref{eq:defLbar} 
\begin{equation}\label{eq:defLa}
\bar L(a,\bar a)= T+\bar a\,S+a\,A \quad\text{with}\quad \frac{a^2-\bar a^2-1}{\bar a}=\frac{\lambda^2-\rho^2-1}{\rho}\,.
\end{equation}
Indeed one can check from \eqref{eq:defK} that we have
\begin{equation}
\text{tr}K(x)\,\mL = T+\bar a(x)\,S 
+a(x)A\equiv \bar L(x)\ \text{ with}\
\begin{cases}\displaystyle
\bar a(x)=\frac{\beta_1(x)\beta_2(x)}{8\rho(x+1)}\,,\\[1.6ex] 
\displaystyle
a(x)=(x+1)\frac{(\lambda+\rho)^2-1}{8\rho} -\frac{(\lambda-\rho)^2-1}{2\rho(x+1)}\,.
\end{cases}
\end{equation}
The relation in \eqref{eq:defLa} follows from the explicit form of $\beta_1(x)$ and $\beta_2(x)$.
Some specific values for $x$ provide interesting sub-cases:
\begin{equation}
\begin{aligned}
&x_{0}=1 \quad &\Rightarrow\quad & \bar L(x_0)=T+\lambda\,A+\rho S\,, \\
&x_{\pm}=-1\pm2\sqrt{\frac{(\lambda-\rho)^2-1}{(\lambda+\rho)^2-1}} \quad &\Rightarrow\quad & \bar L(x_{\pm})=T+\rho_{\pm}\,S\,,\quad
\rho_{\pm}=\bar a(x_{\pm})\,,\\
&x_j=-\frac{3\rho-\lambda-(-1)^j}{\rho+\lambda+(-1)^j}\quad &\Rightarrow\quad & \bar L(x_j)=T+\lambda_j\,A\,,\quad
\lambda_j=a(x_j)\,,\ j=1,2\,.
\end{aligned}
\end{equation}
$\bar L(x_0)$ corresponds to the original $\bar L$ matrix, while $\bar L(x_{\pm})$ leads to the Ragnisco--Bruschi Lax matrix $L(\rho)$. The last case corresponds to a deformation of the original Peakon Lax matrix in a direction orthogonal to the one chosen by Ragnisco--Bruschi.

We now establish the Poisson bracket structure between any two matrices of this family.
From the Poisson brackets \eqref{eq:PB1}--\eqref{eq:PB7}, we obtain
\begin{equation}
\begin{split}
\big\{\bar L_1(a,\bar a)\,,\,\bar L_2(b,\bar b)\big\}&= \frac{\nu}4\,\big[\bar a\Gamma_{21}-a(\Pi-\Pi^t)\,,\,\bar L_2(b,\bar b)\big] 
-\frac{\nu}4\,\big[\,\bar b\Gamma_{12}-b(\Pi-\Pi^t)\,,\,\bar L_1(a,\bar a)\big]
\\[1ex]
&\quad+\frac{\nu}4\,\big[\Pi\,,\,(a\bar b+\bar ab)S_2+( ab+\bar a\bar b-1) A_2\big] 
\\
&\quad+\frac{\nu}4\,\big[\Pi^t\,,\,-(a\bar b-\bar ab)S_2+( \bar a\bar b-ab-1) A_2\big] \,.
\end{split}
\end{equation}
Since $\bar L(a,\bar a)$ and $\bar L(b,\bar b)$ are in the same family, $\frac{a^2-\bar a^2-1}{\bar a}=\frac{b^2-\bar b^2-1}{\bar b}$. 
Then one identifies consistently
\begin{equation}
\begin{cases}
\displaystyle(a\bar b+\bar ab)S+( ab+\bar a\bar b-1) A = c\big(\bar L(a,\bar a)-\bar L(b,\bar b)\big)\\[2ex]
\displaystyle-(a\bar b-\bar ab)S+( \bar a\bar b-ab-1) A = c'\big(\bar L(a,\bar a)^t-\bar L(b,\bar b)\big)
\end{cases}
 \text{where}\ \ 
\begin{cases}
\displaystyle c=\frac{a\bar b+\bar ab}{\bar a-\bar b}\\[2ex]
\displaystyle c'=-\frac{a\bar b-\bar ab}{\bar a-\bar b}
\end{cases}
\end{equation}
It leads to 
\begin{equation}\label{eq:PB-LL}
\begin{split}
\big\{\bar L_1(a,\bar a)\,,\,\bar L_2(b,\bar b)\big\}&= \frac{\nu}4\,\big[\bar a\Gamma_{21}-a(\Pi-\Pi^t)\,,\,\bar L_2(b,\bar b)\big] 
-\frac{\nu}4\,\big[\,\bar b\Gamma_{12}-b(\Pi-\Pi^t)\,,\,\bar L_1(a,\bar a)\big]
\\[1ex]
&\quad+\frac{\nu}4\,c\big[\Pi\,,\,\bar L_2(a,\bar a)-\bar L_2(b,\bar b)\big] 
+\frac{\nu}4\,c'\big[\Pi^t\,,\,\bar L_2(a,\bar a)^t-\bar L_2(b,\bar b)\big] \,.
\end{split}
\end{equation}
Finally, using once more the relations $\big[ \Pi,M_2 \big]=-\big[ \Pi,M_1\big]$ and $\big[ \Pi^t,M_2^t \big]=\big[ \Pi^t,M_1\big]$ (for any matrix $M$), we can rewrite the above relation 
in a $r$-matrix form:
\begin{equation}\label{eq:PB-rLL}
\begin{split}
\big\{\bar L_1(a,\bar a)\,,\,\bar L_2(b,\bar b)\big\}&= \big[\br_{12}(a,\bar a; b,\bar b)\,,\, \bar L_1(a,\bar a)\big] 
- \big[\br_{21}(b,\bar b;a,\bar a)\,,\, \bar L_2(b,\bar b)\big] \\
\br_{12}(a,\bar a; b,\bar b) &= \frac{\nu}4\,\Big(
-\bar b\,\Gamma_{12}+b\,(\Pi-\Pi^t)-c\,\Pi+c'\,\Pi^t \Big)\,,\\
\br_{21}(b,\bar b;a,\bar a) &= \frac{\nu}4\,\Big(-\bar a\,\Gamma_{21}+a\,(\Pi-\Pi^t)
+c\,\Pi +c'\,\Pi^t\Big)\,.
\end{split}
\end{equation}
Using the values of $c$ and $c'$, one can consistently check that $c\big|_{(a,\bar a)\leftrightarrow(b,\bar b)}=-c$ and $c'\big|_{(a,\bar a)\leftrightarrow(b,\bar b)}=c'$, justifying the notation 
$\br_{21}(b,\bar b;a,\bar a)$ and $\br_{12}(a,\bar a;b,\bar b)$ in \eqref{eq:PB-rLL}.

Let us stress that despite $\br$ apparently depends on four spectral parameters, there are only two independent ones on the complex variety $\frac{a^2-\bar a^2-1}{\bar a}=\frac{\lambda^2-\rho^2-1}{\rho}$, as it should be for an $r$-matrix.
Indeed, one can parametrize the variety as
\begin{equation}
\begin{aligned}
&\bar a=\frac1{\alpha\,\sinh(z)-\frac{\gamma}2}\qquad;\qquad
a=\frac{\alpha\,\cosh(z)}{\alpha\,\sinh(z)-\frac{\gamma}2}\,,\quad z\in\CC\\
&\mbox{with}\quad \alpha^2=1-\frac{\gamma^2}4\qquad;\qquad\gamma=\frac{\lambda^2-\rho^2-1}{\rho}\,.
\end{aligned}
\end{equation}
Then, after a rescaling $\bar L(a,\bar a)\,\to\,\frac1{\bar a}\,\bar L(a,\bar a)$, we get an $r$-matrix of the form
\begin{equation}
\begin{split}
 \br_{12}(z_1; z_2)=& \frac{\nu}4 \Big( -\Gamma_{12} 
 +\frac{\sinh(z_1)+\sinh(z_2)}{\cosh(z_1)-\cosh(z_2)}\big(\alpha\sinh(z_2)-\frac{\gamma}2\big)\,\Pi \\
 &+\frac{\sinh(z_1)+\sinh(z_2)}{\cosh(z_1)+\cosh(z_2)}\big(\alpha\sinh(z_2)-\frac{\gamma}2\big)\,\Pi^t
 +\alpha\,\cosh(z_2)\,\big(\Pi-\Pi^t\big) \Big).
\end{split}
\end{equation}
Note also that the existence of a non-dynamical, spectral parameter dependent $r$-matrix is not inconsistent with the fact that the initial Lax matrix $\bar L(\lambda,\rho)$ in \eqref{eq:defLbar} does not possess a $r$-matrix.
Indeed, in this case we have $a=b=\lambda$ and $\bar a=\bar b=\rho$. Then, since $\br(a,\bar a;b,\bar b)$ is singular when $\bar a=\bar b$, 
the corresponding Poisson bracket has to be extracted by a non-trivial limiting procedure, breaking the $r$-matrix form.

\paragraph{Conserved quantities without extended auxiliary space.\label{sec:tau-a}}
Now that we have expressed the Poisson brackets of $\bar L_1(a,\bar a)$ and $\bar L_2(b,\bar b)$ in the $r$-matrix form \eqref{eq:PB-rLL}, standard
arguments of $r$-matrix presentation show that the invariant quantities
\begin{equation}
\tau_n(a,\bar a)=\text{Tr}\big(\bar L(a,\bar a)\big)^n\,
\end{equation}
 PB-commute:
\begin{equation}\label{eq:PB-tau-a}
\big\{\tau_n(a,\bar a)\,,\,\tau_m(b,\bar b)\}=0\,.
\end{equation}
Since $\tau_n\big(a(x),\bar a(x)\big)\equiv\tau_n(x)$, this ends the proof of property \eqref{eq:PBtaux}.

\subsubsection*{Acknowledgments}
J.A. wishes to warmly thank LAPTh for their kind hospitality and financial support.
We wish to warmly thank the referees for their careful reading of the manuscript and their comments, which in particular contribute to the refinement of section 
\ref{sec:spec}.

\end{document}